# Tailoring the defects and electronic band structure in WS$_2$/h-BN heterostructure

Suvodeep Paul, Saheb Karak, Saswata Talukdar, Devesh Negi, and Surajit Saha*

*Department of Physics, Indian Institute of Science Education and Research Bhopal, Bhopal 462066, India.*

(Corresponding author: surajit@iiserb.ac.in)

**Abstract**

The 2D semiconducting transition metal dichalcogenides (*e.g.*, WS$_2$) host strong coupling between various degrees of freedom leading to potential applications in next-generation device applications including optoelectronics. Such applications are strongly influenced by defects which can control both the optical and electronic properties of the material. We demonstrate the possibility to tailor the defect-related electronic states and the lattice dynamics properties of WS$_2$ in their heterostructures with h-BN which host a strong interlayer coupling between the charge carriers in the WS$_2$ layer and the phonons of h-BN. This coupling is observed to induce modifications to the interlayer phonons (manifested by their modified Raman-activity) and to the charge carrier mobilities in the WS$_2$ layer (which results in creation of mid-gap energy states associated with many-body quasiparticle states). Our study also includes a detailed characterization of the defects through Raman measurements revealing an $A_{1g}$-type nature with differential resonance behavior for the modes that are related to defect scattering with respect to the other normal phonon modes of WS$_2$.

**Introduction**

The two-dimensional (2D) layered materials essentially constitute sheets of nanoscale dimensions that are weakly held together by van der Waals interactions, thereby, allowing the separation of these 2D sheets. Over the past decade, scientists have been successful in obtaining 2D sheets of graphene and various other 2D materials which host a plethora of exciting properties [1]. The discovery of graphene and subsequently other 2D materials marked a revolution in the new-generation device technologies [2]. The reduced dimensionality and flexibility associated with the thin flakes of 2D materials can push the limits of nanoscale devices [3]. Additionally, confinement and quantum interactions between various degrees of freedom in 2D materials promise breakthroughs in a variety of novel technologies including electronics [4], optoelectronics [5], spintronics [6], valleytronics [7], magnonics [8], *etc*. Importantly, the performance of the devices fabricated using these 2D materials are often dictated by the presence of defects and disorders [9]. For example, defects can significantly alter the charge mobility [10] in the devices, thereby, affecting their performances. Further, point defects can also change the electronic band structure [11] locally resulting in modifications of the bandgap and direct to indirect bandgap transitions [12], thereby significantly changing the electronic as well as optical properties [13,14]. Therefore, extensive efforts have been put into repair and prevention of defects in 2D materials in order to retain their properties [15,16]. Defects, however, can also have quite useful applications under certain circumstances [17,18]. For example, it has been reported that point defects can result in an overall increase in the light emission from 2D semiconductors [14]. Further, new emission peaks, associated with defect states can be generated, which in turn show promise in the development of multi-colored light emitting devices [14,19–21]. It is, therefore, desirable to obtain better control on the defects to engineer functionalities in the 2D materials-based devices. Van-der Waals heterostructures, which are hybrid structures prepared by stacking various 2D materials, can serve as an important means to achieve such control on the defects present in 2D materials [16]. It has been reported that the effect of extrinsic defects can be reduced to a great degree by heterostructure engineering, thereby, improving the device performances [22]. Further, by choosing suitable materials, it may be possible to induce charge injection required to generate optically active defect states, and therefore, desired quantum emission [16]. Certain van der Waals heterostructures also exhibit interlayer coupling between the various degrees of freedom, which may in turn lead to other functionalities associated with defect states [23–26].

Motivated by these developments, we have chosen 2H-WS$_2$ as a prototypical system to study the characteristics of defects and the possibility to engineer them through the preparation of heterostructures. Notably, 2H-WS$_2$ has been reported to show strong photoluminescence (PL) emissions dominated by the excitonic states, therefore, promising great potential in optoelectronic device applications. We have prepared heterostructures of WS$_2$ with h-BN, which have been previously reported [23] to host interlayer coupling between the phonons of h-BN and the electronic continuum in WS$_2$.

The characterization of defects in 2D materials and van-der Waals heterostructures rely mainly on optical techniques like Raman and PL studies as they are non-destructive and easy to perform. Generally, signatures of defects in Raman measurements are observed as defect-induced modes due to second- (or higher-) order processes where the phonons corresponding to a non-zero momentum point of the Brillouin zone are elastically scattered by defects (resulting in satisfaction of Raman selection rules) [27,28]. While the PL features corresponding to point-defects, *viz.*, defect-bound excitons, biexcitons, etc. have been very well explored for $WS_2$ [20,21], the corresponding Raman signatures have not been sufficiently studied. For $2H$-$MoS_2$, Raman measurements have predicted the $LA(M)$ mode to be a defect-generated mode that shows an increase in intensity with increasing concentration of defects [28]. Unfortunately, no such studies are reported for $2H$-$WS_2$. There are certain theoretical [29] and a tip-enhanced Raman spectroscopy (TERS) measurement [30] that report an out-of-plane phonon to be associated with defect scattering in $WS_2$. However, lack of ample number of experimental evidences calls for further investigations. Additionally, inelastic light scattering has been recently used to detect strong magneto-optic effects in TMDs on application of a magnetic field [31,32]. The magneto-optic effect was also reported to be sensitive towards the presence of defects in the crystal [33]. Though this was supported by probing the first-order out-of-plane $A_{1g}$ phonons, it would be naturally interesting to observe such effects of magnetic field in the defect-induced modes of TMDs.

In this paper, we have investigated flakes of $2H$-$WS_2$ and their heterostructure with h-BN through micro-Raman and PL measurements. We have observed enhancement (or appearance) of certain Raman modes in the heterostructure, which when complemented by the PL studies could be assigned to defect scattering. The presence of additional features in the PL spectrum were observed at low temperatures which could be associated with defect-bound excitonic and biexcitonic states. The temperature-dependent Raman measurements also showed differential resonance behaviors for the Raman modes, where the defect-generated modes resonated with the defect states, but all other first-order and second-order Raman modes resonated with the excitonic states. Finally, our magneto-Raman studies also revealed a strong magneto-optic modulation of the defect-originated Raman modes, in addition to the first-order out-of-plane phonons.

**Experimental details**

The WS$_2$ samples were micromechanically exfoliated using a scotch tape and transferred on to a Si/SiO$_2$ substrate. Further, in order to prepare the heterostructure, a h-BN flake was similarly exfoliated and transferred on to a transparent PDMS film. Subsequently, the PDMS film was positioned over the pre-transferred WS$_2$ flake on the SiO$_2$/Si substrate using a microscope in a custom-built micromanipulator. The h-BN layer was then transferred on top of the WS$_2$ layer by a dry transfer technique to obtain the desired heterostructure. The flake thicknesses of the WS$_2$ and h-BN layers were measured by atomic force microscopy measurements performed using an Agilent 5500 system in a non-contact mode. The Raman and photoluminescence measurements were performed using a Horiba JY LabRam HR Evolution Raman spectrometer fitted with a 50× (NA=0.5) long-working distance objective lens. The scattered light was collected in a back-scattering geometry, passed through a grating of 1800 grooves/mm and finally detected using an air-cooled charge coupled device detector. In order to study the resonance effects with the A and B excitons of WS$_2$, the samples were excited by a 633 nm (He-Ne gas laser) and a 532 nm (Nd:YAG-diode laser), respectively. The sample temperature (from 80 to 300 K) was varied using a LINKAM cryostat (Model No. HFS600EPB4). The low temperature (5 K) and magnetic-field dependent Raman measurements were performed in a closed cycle Helium AttoDRY 1000 cryostat.

**Results and Discussion**

The Raman and PL measurements performed on the uncapped WS$_2$ and the WS$_2$-hBN heterostructure bring out certain exciting observations which have been elaborated in this section. The results have been divided into the following sections for the benefit of the reader.

1. **Atomic structure and optical characterization**

The transition metal dichalcogenides (TMDs) are represented by the molecular formula of MX$_2$ and constitute a layer of transition metal (M) atoms sandwiched between two layers of chalcogen (X) atoms. Hexagonal BN has a planar structure with alternate B and N atoms arranged in a honeycomb lattice. Fig. 1a shows a composite structure (heterostructure) of 2H-WS$_2$ and h-BN, with h-BN acting as a capping layer for WS$_2$. The empty space between the 2H-WS$_2$ and h-BN layers represents the van der Waals gap. Fig. 1b (top panel) compares the Raman spectra of the uncapped WS$_2$ and the WS$_2$-hBN heterostructure (corresponding optical images are shown in Fig 1d) obtained using the 532 nm laser excitation line. The prominent modes observed are labelled from $P_1$-$P_{10}$. The corresponding phonon assignments, based on prior reports [34–36], have been enlisted in Table 1. Notably, in addition to the first-order $\Gamma$-point phonons, $E_{2g}^1(\Gamma)$ (labelled as $P_9$) and $A_{1g}(\Gamma)$ (labelled as $P_{10}$), we observe several second-order modes, which appear as a consequence of the resonance with the B exciton bandgap. The heterostructure of $WS_2$-hBN has a very similar Raman signal to the uncapped $WS_2$ flake, with some prominent differences towards the low frequency region, *viz.*, new modes ($P_1$ and $P_2$) appear while the mode $P_3$ is strongly enhanced. The heterostructure also shows a Raman mode at ~1366 cm$^{-1}$, representing the $E_{2g}$ phonon corresponding to the in-plane B-N vibrations of the h-BN capping layer [37,38] (refer to Supplementary figure SF1 [39]). All these features remain intact in the Raman spectra obtained using the 633 nm laser excitation (Fig. 1b (bottom panel)), where the resonance occurs with the A exciton bandgap.

**Table 1**

| Label | Phonon assignment | Frequency (cm$^{-1}$) |
|---|---|---|
| $P_1$ | $LA(K) - 2E_{2g}^2(\Gamma)$ | 140 |
| $P_2$ | $LA(M) - E_{2g}^2(\Gamma)$ | 148 |
| $P_3$ | $LA(M)$ | 176 |
| $P_4$ | $LA(K)$ | 190 |
| $P_5$ | $A_{1g}(M) - LA(M)$ | 230 |
| $P_6$ | $2LA(M) - 2E_{2g}^2(\Gamma)$ | 297 |
| $P_7$ | $2LA(M) - E_{2g}^2(\Gamma)$ | 324 |
| $P_8$ | $2LA(M)$ | 351 |
| $P_9$ | $E_{2g}^1(\Gamma)$ | 355 |
| $P_{10}$ | $A_{1g}(\Gamma)$ | 417 |

In order to determine the origin of the new modes ($P_1$ and $P_2$) and the enhancement of the $P_3$ mode, we may note from Table 1 that the phonon assignment corresponding to the $P_3$ mode is $LA(M)$. The $LA(M)$ phonon has been attributed to a defect-induced mode for various

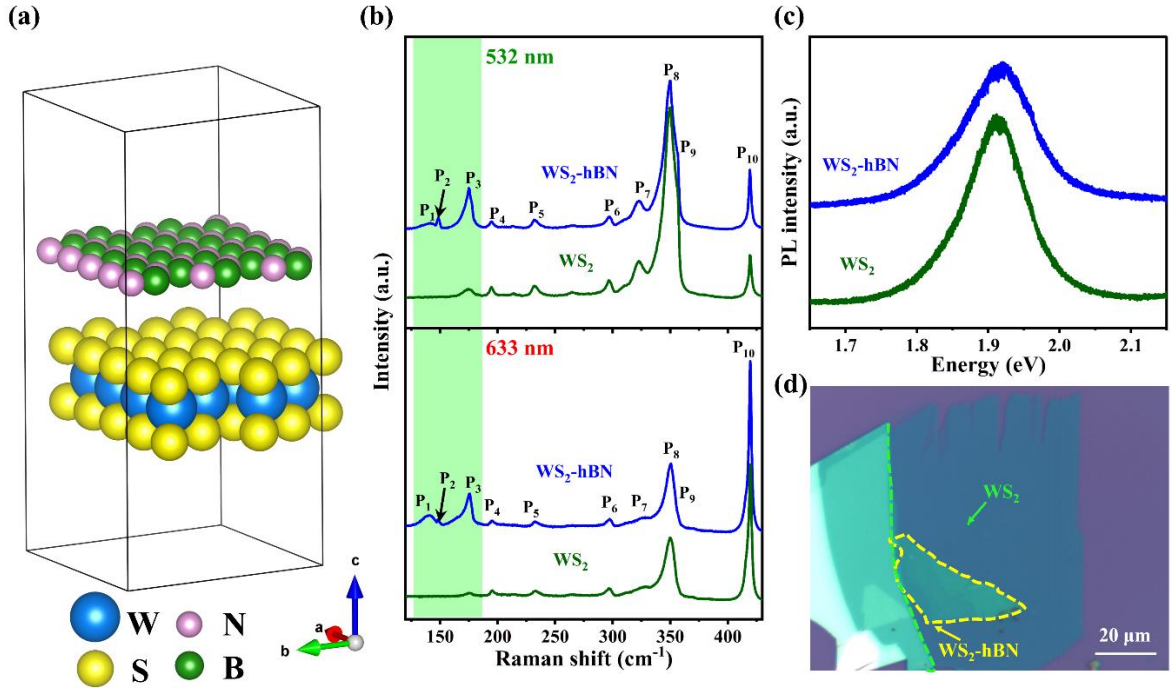

**Figure 1.** (a) The structure of the $WS_2$-hBN heterostructure. (b) Raman spectra of the uncapped $WS_2$ and the $WS_2$-hBN heterostructure using the 532 nm (top panel) and 633 nm (bottom panel) excitation laser lines. The part enclosed in the green shaded region shows the enhancement of the mode $P_3$ and the appearance of new modes $P_1$ and $P_2$ in the $WS_2$-hBN heterostructure as compared to uncapped $WS_2$. (c) Photoluminescence spectra of the investigated samples. (d) The optical microscope images showing the investigated samples. The green dashed line shows the edge of the $WS_2$ flake and the yellow dashed line marks of the h-BN capping layer

TMDs [28,34]. Conventionally, the Raman activity of a phonon is determined by its symmetry. The most general selection rule imposed on a Raman scattering process to generate Raman-active phonons is to satisfy the momentum conservation selection rule, which may be written as: $\boldsymbol{k_s} = \boldsymbol{k_i} \pm \boldsymbol{q}$, where the $\boldsymbol{k_s}$, $\boldsymbol{k_i}$, and $\boldsymbol{q}$ vectors represent the momenta of the scattered photon, the incident photon, and the phonon, respectively. Due to the negligible length of interatomic distances in a crystal when compared to the wavelength of the incident and scattered photons, we can readily infer that the momentum conservation selection rule, effectively reduces to $\boldsymbol{q} \approx \boldsymbol{0}$ (corresponding to near the zone-center or near the $\Gamma$-point of the Brillouin zone) for a phonon to be Raman active. The introduction of defects in a crystal, however, breaks the translational symmetry of the crystal, thereby, resulting in a relaxation of the momentum-conservation selection rule. This leads to activation of defect-induced modes which essentially can originate from non-zero momentum points (the zone-boundary or $M$-point of the Brillouin zone, for instance), when the elastic scattering from the defects in the crystal provides the required momentum to satisfy the momentum conservation. The intensity of these defect-induced modes depends on the concentration of defects in the crystal, and may be used to quantify defect density. In fact, Mignuzzi *et al.* [28] reported that the intensity of the $LA(M)$ mode can be used to quantify the concentration of defects in the MoS$_2$ crystals. Unfortunately, the presence of the $LA(M)$ phonon in the Raman spectra obtained for pristine samples of WS$_2$, has led to an ambiguity about its association with defects [30]. In this regard, we may note that the WS$_2$ crystals show strong resonance phenomena with the A and B excitonic bandgaps when excited by the commonly used laser excitations of 633 nm and 532 nm, respectively. As a result, the second-order Raman scattering processes, including the defect-induced modes, may show strong enhancement. We believe that this leads to the appearance of the P$_3$ mode ($LA(M)$ phonon) in the pristine samples of WS$_2$, as it is impossible to get a truly pristine (defect-free) sample when exfoliated under ambient conditions. Additionally, we may also note that the modes P$_1$ and P$_2$, which appear only in the heterostructure, are also combination modes related to the $LA$ phonon (refer to Table 1). We may also note from the polarization-dependent measurements (Supplementary figure SF3 [39]) that the P$_1$, P$_2$, and P$_3$ modes show exactly similar dependence on polarization angle. We may, therefore, infer that the enhancement of the P$_3$ mode and the appearance of the P$_1$ and P$_2$ modes in the WS$_2$-hBN heterostructure are all consequences of greater concentration of defects in the heterostructure. Further confirmations of the association with defects will be discussed in subsequent sections.

While the association of the $LA(M)$ mode with defects in WS$_2$ was not well established previously, Lee *et al.* [30] reported a defect-related Raman mode (D), which appears as a shoulder peak of P$_{10}$ ($A_{1g}(\Gamma)$ phonon) through TERS measurements. Theoretical quantum mechanical simulations performed on model WS$_2$ crystals with sulfur vacancies revealed the D mode to have originated from $A'_1(k)$ phonon vibrations [29]. The room-temperature Raman spectra shown in Fig. 1b and Fig. 1c do not resolve this D mode clearly, because of its proximity to the P$_{10}$ mode accompanied by the strong effect of thermal broadening. However, the measurements performed at 80 K distinctly show the D mode (refer to Supplementary figure SF4 [39]). As expected from the trends in the defect-related P$_1$, P$_2$, and P$_3$ modes, we again observe a stronger D mode in the WS$_2$-hBN heterostructure as compared to the uncapped WS$_2$ flake.

Fig. 1d shows the PL spectra for the uncapped $WS_2$ flake and the $WS_2$-hBN heterostructure obtained at room temperature. The PL features for both the samples are dominated by the A exciton transition [40]. However, the intensity of the PL feature is weaker in the heterostructure as compared to the uncapped $WS_2$. This again may be related to dominance of defects in the heterostructure as defects are known to act as recombination centers for electron-hole pairs, thereby resulting in a decrease of the PL intensity [41]. The PL signal for the heterostructure also appears to be broader than the uncapped $WS_2$. Such broadening of the exciton peaks is observed as a consequence of defects in the crystal [21]. Therefore, the Raman and PL measurements at room temperature provide ample evidences that the heterostructure of $WS_2$-hBN host more defect related phenomena relative to the uncapped $WS_2$ flake.

## 2. Electronic transitions corresponding to defects

The room temperature PL measurements discussed in the previous section already indicate the greater influence of defects in the $WS_2$-hBN heterostructure. It is known from various theoretical and experimental studies that the presence of defects in the semiconducting TMDs creates defect states within the excitonic bandgap [19,41–45]. Electronic transitions corresponding to these defect-related mid-gap states result in additional features in the PL spectrum below the exciton ($A_0$) and trion ($A_-$) features. Fig. 2a shows contour plots comparing the PL signatures of the uncapped $WS_2$ and the $WS_2$-hBN heterostructure obtained at 80 K as a function of excitation laser power. We observe that the PL is largely dominated by the contributions from the A exciton ($A_0$) / trions ($A_-$), centred at ~2.03 eV. Additionally, we observe certain distinct differences in the signatures of the two samples. First, we may note that the PL emission is stronger in the uncapped $WS_2$ compared to the heterostructure, as observed from the red region in the contour plot (also see the stacks of PL spectra shown in Supplementary figure SF5 [39]). This is typically expected in presence of defects as the defects may act as recombination centers for the electron-hole pairs, therefore, destroying the excitons [41]. Further, we also observe certain additional features below the $A_0$ or $A_-$, which appear as an asymmetric growth of intensity below the exciton energy, particularly for the higher laser excitation powers. These additional features mainly comprise of defect-bound excitons (D′) [20] or biexcitons (AA) [21]. It has been previously reported that strong D′ and AA features are generally observed in the vicinity of point defects and in the presence of large concentration of charge carriers [21]. The PL features obtained have been deconvoluted into the component peaks corresponding to the various electronic states in the mid-gap region of the electronic band structure (refer to Supplementary figure SF6 [39]). The intensity of the obtained features after deconvolution are plotted as a function of excitation laser power (Fig. 2b) for the heterostructure of $WS_2$-hBN and fitted with a power law expression $I \propto P^m$ (where $I$, $P$, and $m$ represent the intensity, excitation laser power, and numeric power, respectively). It is well-known that the $A_0$ feature shows a linear ($m = 1$) dependence on the laser excitation power, while the D′ and AA features, associated with defects show sub-linear ($m < 1$) and super-linear ($m > 1$) behaviors, respectively [20,21]. Our observation of all these features clearly indicates the presence of defects and the consequential creation of biexciton and defect-bound exciton states below the exciton state in the electronic band structure.

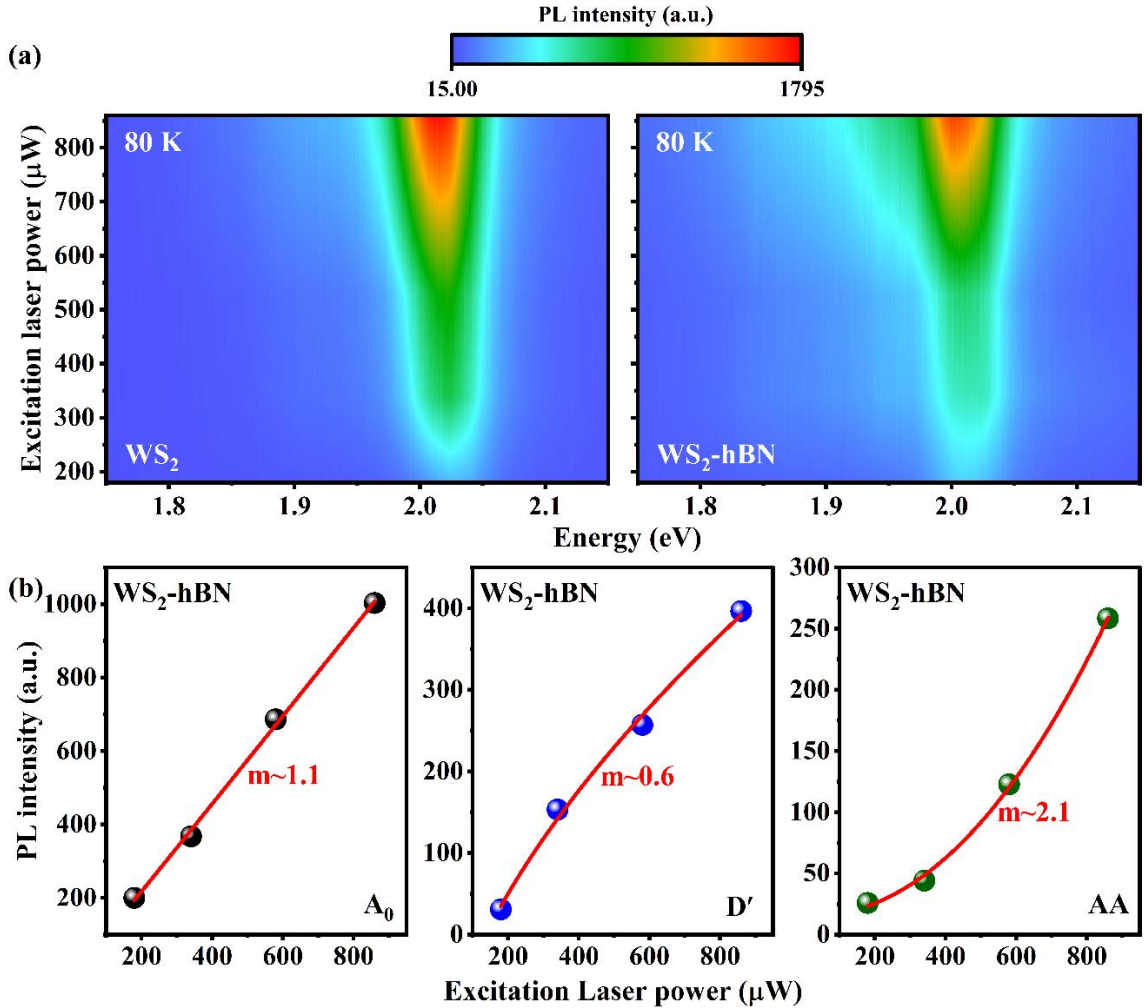

**Figure 2.** (a) The intensity profiles of the PL spectra obtained at 80 K with varying excitation laser power for the uncapped $WS_2$ (left panel) and the $WS_2$-hBN heterostructure (right panel). (b) The intensities corresponding to the exciton ($A_0$), the defect-bound exciton (D′), and the biexciton (AA) obtained from the PL spectra of $WS_2$-hBN heterostructure are plotted as a function of excitation laser power showing linear (m~1.1), sublinear (m~0.6), and superlinear (m~2.1) behaviors, respectively.

## 3. Differential resonance behaviors showcased by the defect-related modes

We have performed extensive temperature-dependent Raman studies on the $WS_2$ and $WS_2$-hBN heterostructure samples. The effect of temperature on phonons is customarily reported as the manifestation of anharmonic behaviors, *viz.*, redshifts of phonon frequencies and broadening of phonon linewidths as a function of temperature [35,46]. However, the dependence of the phonon intensities on temperature is particularly important in the study of resonance phenomena [47]. Resonance Raman scattering phenomena can occur when the incident photon possesses an energy which is equivalent to an electronic transition, resulting in an enhancement of the Raman signal by several orders of magnitude. The intensity of the Raman signal is proportional to the closeness of the photon's energy to the electronic transition. A detailed study of resonance Raman may, therefore, be performed by using various laser excitation sources [48,49], which may often be inconvenient or expensive. Alternatively, varying the temperature of the sample can also induce changes in the electronic band structure, thereby, driving the sample through the resonance condition, albeit when the laser excitation source is nearly equivalent to the electronic transitions available in the system. Our temperature-dependent Raman measurements reveal the resonance effect very clearly. Fig. 3 shows the intensity profiles for various Raman modes (of the $WS_2$-hBN heterostructure) as a function of temperature. We notice that the intensity grows with temperature initially and finally drops again. The corresponding maxima in the intensity profiles (the most intense red portions of the contour plots in Fig. 3a) represents the best resonance conditions for the modes. The first order Raman modes ($P_9$ and $P_{10}$), and the second-order $2LA(M)$ mode ($P_8$) have their maxima at ~280 K, as also separately shown in the plots in Fig. 3b. The behavior is consistent with various experimental reports on resonance Raman measurements of $WS_2$ flakes. It is worth mentioning here that due to the close proximity of the $P_8$ and $P_9$ mode frequencies, we could not resolve them, and hence the observed resonance behavior is seen as a result of the combination of the two modes. However, as both of these modes show a very similar trend in the resonance behavior, the lack of resolution does not suppress the actual behavior of either mode. The maxima for the various modes mentioned above (*viz.*, P$_8$, P$_9$, and P$_{10}$) are reached due to resonance with the A exciton bandgap, which decreases as a function of temperature. This can be further confirmed from the temperature-dependent PL measurements, which clearly show a redshift in the PL feature corresponding to the emissions from the A exciton, as shown in Supplementary figure SF7 [39]. Interestingly, we observe that the modes P$_1$, P$_2$, P$_3$, and D show a similar resonance behavior, but with their maxima located at a relatively lower temperature (~240 K). This leads us to conclude that the resonance condition met by the modes P$_1$, P$_2$, P$_3$, and D are associated with different electronic transitions than the other modes. In this regard, we can recall from our discussions in the previous sections that the modes $P_1$, $P_2$, $P_3$, and D are all associated with defect-scattering and that defects result in the creation of new-mid gap states related to defect-bound excitons and biexcitons in the electronic band structure. Therefore, the phonon modes that are originated by defect-scattering processes may show a resonance with the defect-related electronic transitions. Now, as the increment of temperature is responsible for the reduction of bandgap in semiconductors, it is likely that while driving the sample temperature from low to high, the resonance condition would be met first for the defect related electronic levels (which occur at a lower energy) compared to the exciton level. The situation is schematically explained in Fig.4. We observe that at temperatures T<240 K, the resonance condition is not met for any of the states. As the sample is heated to T=240 K, the laser excitation of 1.96 eV resonates with the defect-induced electronic state D′. The Raman-active modes that show resonance with this defect-induced level ($P_1$, $P_2$, $P_3$, and D) should reach their maxima at this temperature. Further, as the temperature is increased, these modes again go off-resonance, until at T=280 K, the resonance condition is met again but for the exciton state, where all the Raman modes (excepting the ones related to defect scattering) reach their maxima. The above discussion and the Figs. 3 and 4 are related to the data collected using the 1.96 eV laser line and all resonance phenomena occur with the A exciton ($A_0$) and the corresponding defect-states. Similarly, when the experiments are performed with 2.33 eV, similar behaviors would appear as a result of resonance with B excitons ($B_0$) and corresponding defect-bound excitons or biexcitons associated with the B exciton. The Raman measurements for the 2.33 eV excitation are shown in Supplementary figure SF8 [39], exhibiting the resonance phenomena explained above.

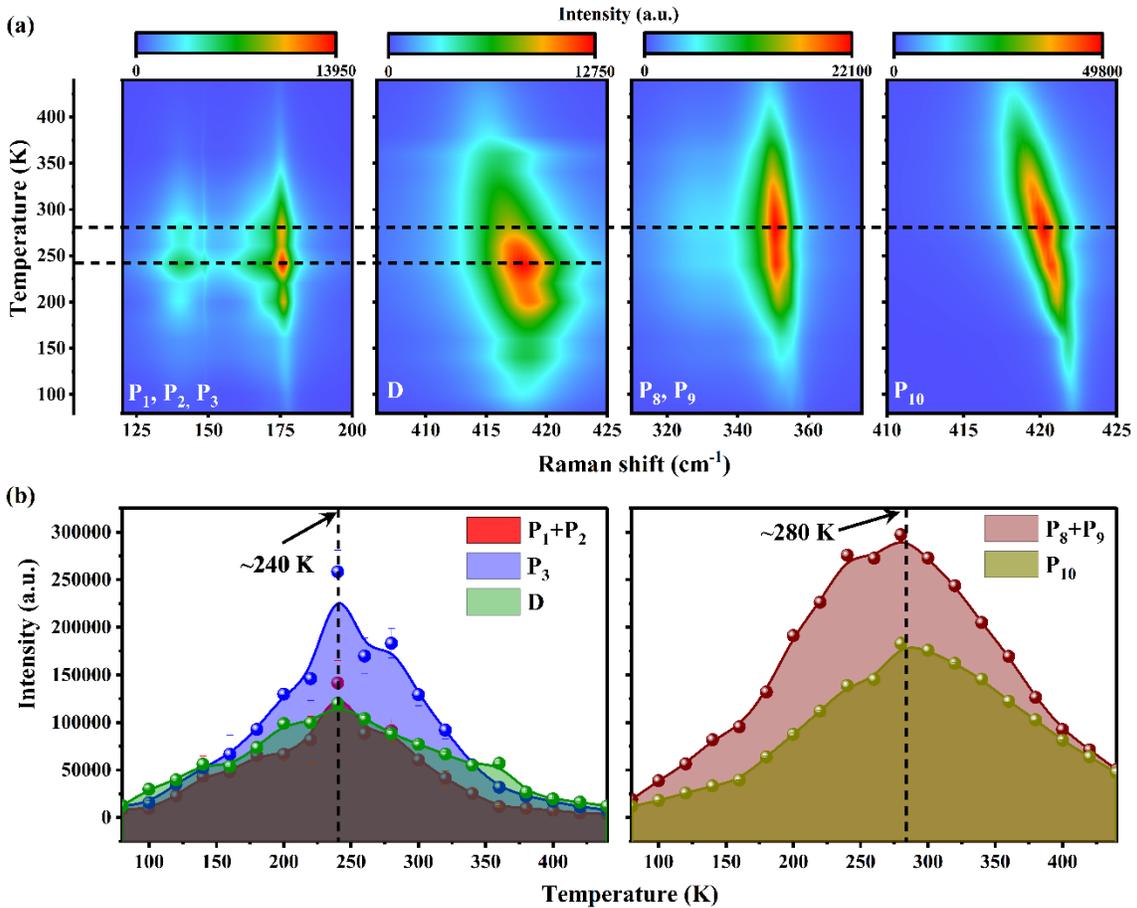

**Figure 3.** **(a)** The intensity profiles corresponding to the various prominent Raman modes of the WS$_2$-hBN heterostructure are shown. The maximum intensity for the various modes corresponds to the resonance condition. **(b)** The intensity of the defect-related modes (left) and the other prominent Raman modes unrelated to defects (right) clearly show the resonance phenomena at different temperatures.

## 4. Magneto-optic effect observed in defect-induced modes

The application of magnetic field on TMDs like $MoS_2$ have been reported to cause a giant magneto-optic effect where the $A_{1g}(\Gamma)$ phonon modes involving the out-of-plane vibrations of the chalcogen (S) atoms show a field-dependent variation in intensity [31]. The effect has been attributed to the Lorentzian forces acting on the electrons and a consequent reduction of the crystal symmetry in presence of the magnetic field that modifies the Raman scattering process [31,32]. Wan *et al.* [32] demonstrated that the effect is a result of the rotation of the polarization plane of the Raman scattered photon associated with the creation of the out-of-plane $A_{1g}(\Gamma)$ phonon in presence of the perpendicular magnetic field. Such an effect was not observed for the in-plane $E_{2g}^1(\Gamma)$ phonons [32]. In this regard, we may also note that the in-plane $E_{2g}^1(\Gamma)$ phonons show no polarization dependence, i.e., the Raman cross-section for the $E_{2g}^1(\Gamma)$ phonons is not a function of the polarization angle. Therefore, a rotation of the polarization plane of the photon induced by the magneto-optic effect, intuitively should not have any impact on the intensity of the in-plane $E_{2g}^1(\Gamma)$ phonon [46]. We have performed magnetic-field dependent experiments on the uncapped WS$_2$ and the WS$_2$-hBN heterostructure. Fig. 5 shows the magnetic-field dependent Raman signals for the two samples obtained in XX (parallel) and XY (perpendicular) polarization configurations at 5 K, using the 532 nm laser line as the excitation source. Our results are consistent with the various reports of magneto-optic effect observed in MoS$_2$ [31,32]. We observe that while the $E_{2g}^1(\Gamma)$ phonon (P$_9$) shows no appreciable variation in intensity as a function of magnetic field, the $A_{1g}(\Gamma)$ phonon (P$_{10}$) and the defect-related mode D (which also has been reported to have $A_1'$ symmetry) clearly show a strong variation in intensity. The D mode is observed to be stronger in the WS$_2$-hBN heterostructure than the uncapped WS$_2$, as also observed in the previous sections. It is known that the out-of-plane modes having *A*-type symmetry are Raman active in the parallel configuration and are suppressed in the perpendicular configuration by virtue of the symmetry-based Raman selection rules. This is clear as in absence of the magnetic field (refer to 0 T data), we observe the modes P$_{10}$ and D to be present in the Raman signals obtained in the XX configuration and absent in the XY. However, as the magnetic field is increased, we observe the intensity to decrease (increase) initially and then increase (decrease) again with a minimum (maximum) at ~5 T for the parallel (perpendicular) polarization spectra.

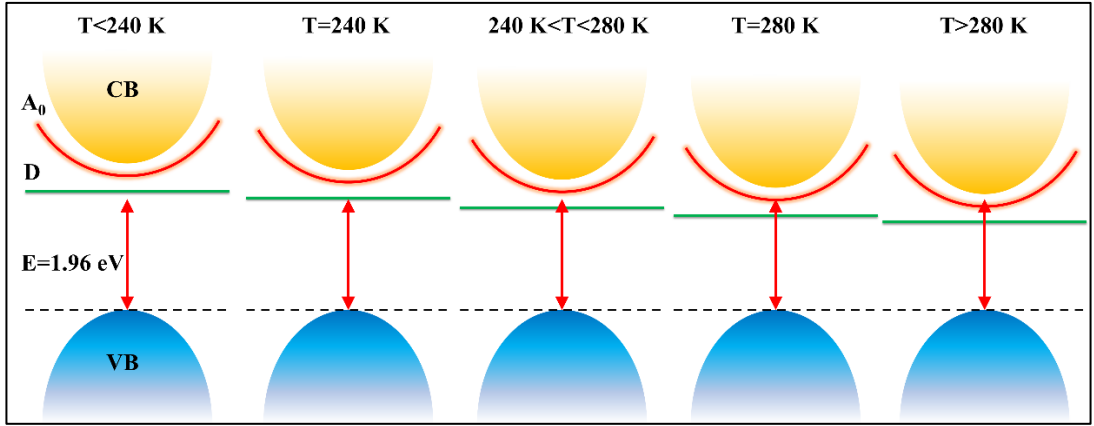

**Figure 4.** Schematic representation of the resonance condition met by the defect-scattered Raman modes and the other Raman modes unrelated to defects, explaining the differential resonance conditions for them.

In addition to this expected behavior for the first-order modes P$_9$ and P$_{10}$, we have also observed interesting results for the defect-related modes P$_1$, P$_2$, and P$_3$ and the second-order P$_8$ mode, which are all related to $LA(M)$ vibrations. (The mode P$_1$ has not been previously reported and is clearly a defect-scattered mode. The phonon assignment for this mode, based on the comparison of the phonon dispersion calculations of WS$_2$ by various groups, however, shows that it may be related to the $LA(K)$ phonons). The longitudinal acoustic phonons are collective movements of the atoms in a crystal that result in periodic compressions and rarefactions along the direction of propagation of the wave. Though the magneto-optic effect is expected only for out-of-plane vibrations, as explained above, we observe very strong magnetic-field dependent intensity modulation of the modes P$_1$, P$_2$, P$_3$, and P$_8$. In order to account for this behavior, we must refer to the eigenvectors corresponding to the $LA(M)$ phonons in TMDs. Bae *et al.* [50] recently performed extensive studies on the longitudinal acoustic phonons of TMDs through theoretical calculations and pump-probe spectroscopy measurements. They have shown that while the $LA$ phonons of the various TMDs hardly show any dispersion in energy between the $K$ and $M$ points of the Brillouin zone, the corresponding eigenvectors reveal very distinct changes. The $LA$ phonons at the $K$ point are associated with in-plane asymmetric vibrations of all the atoms, but the corresponding phonons at the $M$ point show symmetric vibrations with the chalcogen atoms vibrating out-of-plane. As the defect-originated modes (P$_1$, P$_2$, and P$_3$) and the second-order P$_8$ mode are all associated with the $LA(M)$ phonon, we may, therefore, expect to see a similar polarization angle dependence for these modes as seen in the P$_{10}$ ($A_{1g}(\Gamma)$) phonon, also involving out-of-plane vibrations of chalcogen atoms.

## 5. Engineering the defect states in WS$_2$-h-BN heterostructure

Through various Raman and PL measurements, we have observed the predominance of defects in the WS$_2$-h-BN heterostructure. While we could connect the observed behaviors to a greater influence of defects in the WS$_2$-h-BN heterostructure than the uncapped WS$_2$ sample, the possible reasons for the association of the heterostructure with defects has not been discussed so far. Defects in 2D materials may be intrinsic or extrinsic. The intrinsic sources of defects mainly comprise of atomic imperfections and vacancies. While the CVD-grown samples are often reported to show intrinsic defects [21,51,52], the exfoliated flakes are generally considered to be more pristine [14]. Besides, the $WS_2$-hBN heterostructure constitute the same $WS_2$ flake as the uncapped $WS_2$. Therefore, it is unlikely that the heterostructure would possess more intrinsic defects. The other possibility is the presence of extrinsic defects, which refer to the effect of surface adsorbates, surface roughness, charged impurities in the substrate, and oxidation [16]. The capping layer of h-BN is generally believed to protect the 2D flake from the extrinsic defects [22]. For example, the capping layer protects the underlying $WS_2$ flake from any exposure to air and moisture, thereby preventing the oxidation of the surface. Importantly, the extrinsic defects may have come from atomic defects in the capping layer of h-BN itself. However, Tran *et al.* [53] reported that the point defects in h-BN act as quantum emitters from the mid-gap optically active defect states with the strongest and sharpest feature around ~1.99 eV, which is clearly absent in our PL measurements. Therefore, we may rule out most of the possible sources of extrinsic defects as well. We, therefore, propose that the defect related phenomena in our samples appear as a consequence of the proximity to the edge (marked with green dashed line in Fig. 1d) of the $WS_2$ layer in the heterostructure. It has been reported that the PL emission is greatly enhanced at the edges of a TMD layer with strong contributions from the biexcitonic and defect-bound excitonic features [20,21], thereby, supporting our observations. This may be further verified from the absence of the $P_1$, $P_2$, and $P_3$ modes in another heterostructure based on the same $WS_2$ flake (with a capping layer of h-BN) but located away from the edge (refer to Supplementary figure SF9 [39]). Further, we may also consider another possibility here. It has been reported that strong PL emissions corresponding to the defect-related electronic states (like the defect-bound excitons and biexcitons) are observed at the vicinity of regions that are rich in charge carriers [21]. Lin *et al.* [23] reported an interlayer electron-phonon coupling (EPC) in the heterostructures of $WS_2$ and h-BN where the electrons/holes in the $WS_2$ layer couple with the phonons in the h-BN layer. Such an interlayer EPC [23–26] is expected to significantly modulate the behavior of the phonons and the charge carriers involved. The effect on the phonons is manifested by the appearance of Raman inactive interlayer modes of h-BN. We have indeed observed similar signatures of low frequency interlayer Raman-inactive phonons of h-BN along with the enhancement of certain Raman active interlayer modes of $WS_2$ in the $WS_2$-h-BN heterostructure (refer to Supplementary figure SF10 [39]). On the other hand, the effect of interactions between the electrons of graphene with the phonons of the underlying substrate has been reported to limit the electron mobility of graphene [54,55].

Analogously, we may assume that the interlayer EPC may induce a degree of localization on the charge carriers of $WS_2$ in the heterostructure, thereby fetching more charge carriers required for the formation of quasiparticles like biexcitons and defect-bound excitons that were observed in the PL emission spectra.

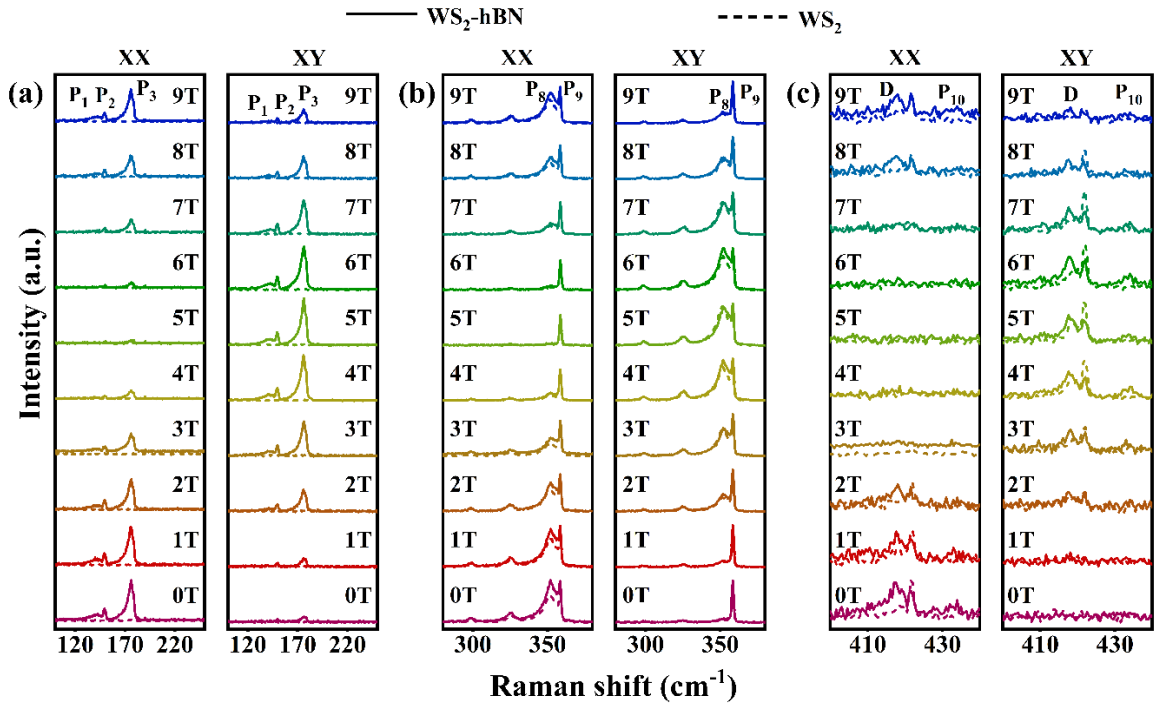

**Figure 5.** The magnetic field dependent evolution of the Raman signatures of the uncapped WS2 (dotted lines) and the WS2-hBN heterostructure (solid lines). The spectra have been acquired in the parallel (XX) and perpendicular (XY) polarization configurations. The spectra have been divided into three parts: **(a)** low-frequency, **(b)** intermediate-frequency, and **(c)** high-frequency regions.

**Conclusion**

In summary, we have studied the defect-related phonon modes in uncapped $WS_2$ and the heterostructure of $WS_2$ and h-BN where we have observed a predominance of the defects in the heterostructure. Our extensive Raman measurements allowed us to assign certain newly observed phonon modes of $WS_2$ to defect scattering phenomena. By means of our temperature-dependent resonance Raman studies, we also observed that defect-related Raman modes show resonance at a comparatively lower temperature than the resonance temperature of all other first-order and second-order Raman modes. This behavior could be explained by our PL studies that show the presence of mid-gap states due to defects below the excitonic bandgap. While the first-order and second-order Raman modes showed resonance with the excitonic energy level, the defect-scattered

Raman phonons showed the resonance behavior with the mid-gap defect-related states. We performed magnetic-field dependent Raman measurements to observe a strong magneto-optic effect which shows a magnetic-field dependent modulation in the intensity of out-of-plane phonons. While such a magneto-optic effect was previously observed in the $\Gamma$-point $A_{1g}$ phonons of $MoS_2$, we observed that the defect-scattered modes associated with $LA(M)$ phonons in $WS_2$ show a similar effect. This could be attributed to the out-of-plane vibration of the chalcogen atoms in the longitudinal acoustic phonons at the $M$-point of the Brillouin zone. Finally, the predominance of defects in the $WS_2$-hBN heterostructure was attributed to the proximity to an edge and the localization of charge carriers induced by the cross-planar electron-phonon coupling in the heterostructure.

**Acknowledgement**

The authors acknowledge funding from the Department of Science and Technology, Science and Engineering Research Board (Grants No. ECR/2016/001376 and No. CRG/2019/002668), Ministry of Education (Grant No. STARS/APR2019/PS/662/FS), Nanomission [Grant No. SR/NM/NS-84/2016(C)], and DST-FIST [Project No. SR/FST/PSI-195/2014(C)]. D.N. acknowledges CSIR for fellowship (09/1020(0139)/2018-EMR-I).

**Competing financial interests:** The authors declare no competing financial interests.

**References**

[1] Z. Xiong, L. Zhong, H. Wang, and X. Li, *Structural Defects, Mechanical Behaviors, and Properties of Two-Dimensional Materials*, Materials (Basel). **14**, 1192 (2021).

[2] R. Roldán, L. Chirolli, E. Prada, J. A. Silva-Guillén, P. San-Jose, and F. Guinea, *Theory of 2D Crystals: Graphene and Beyond*, Chem. Soc. Rev. **46**, 4387 (2017).

[3] H.-Q. Wu, C.-Y. Linghu, H.-M. Lu, and H. Qian, *Graphene Applications in Electronic and Optoelectronic Devices and Circuits*, Chinese Phys. B **22**, 98106 (2013).

[4] S. Manzeli, D. Ovchinnikov, D. Pasquier, O. V Yazyev, and A. Kis, *2D Transition Metal Dichalcogenides*, Nat. Rev. Mater. **2**, 17033 (2017).

[5] J. An, X. Zhao, Y. Zhang, M. Liu, J. Yuan, X. Sun, Z. Zhang, B. Wang, S. Li, and D. Li, *Perspectives of 2D Materials for Optoelectronic Integration*, Adv. Funct. Mater. **32**,

2110119 (2022).

[6]  E. C. Ahn, *2D Materials for Spintronic Devices*, Npj 2D Mater. Appl. **4**, 17 (2020).

[7]  J. R. Schaibley, H. Yu, G. Clark, P. Rivera, J. S. Ross, K. L. Seyler, W. Yao, and X. Xu, *Valleytronics in 2D Materials*, Nat. Rev. Mater. **1**, 16055 (2016).

[8]  R. M. Menezes, D. Šabani, C. Bacaksiz, C. C. de Souza Silva, and M. V Milošević, *Tailoring High-Frequency Magnonics in Monolayer Chromium Trihalides*, 2D Mater. **9**, 25021 (2022).

[9]  Z. Lin, B. R. Carvalho, E. Kahn, R. Lv, R. Rao, H. Terrones, M. A. Pimenta, and M. Terrones, *Defect Engineering of Two-Dimensional Transition Metal Dichalcogenides*, 2D Materials **3**, 022002 (2016).

[10] M. Ghorbani-Asl, A. N. Enyashin, A. Kuc, G. Seifert, and T. Heine, *Defect-Induced Conductivity Anisotropy in MoS 2 Monolayers*, Phys. Rev. B **88**, 245440 (2013).

[11] S. Yuan, R. Roldán, M. I. Katsnelson, and F. Guinea, *Effect of Point Defects on the Optical and Transport Properties of MoS 2 and WS 2*, Phys. Rev. B **90**, 41402 (2014).

[12] B. G. Shin, G. H. Han, S. J. Yun, H. M. Oh, J. J. Bae, Y. J. Song, C. Park, and Y. H. Lee, *Indirect Bandgap Puddles in Monolayer MoS2 by Substrate-induced Local Strain*, Adv. Mater. **28**, 9378 (2016).

[13] R. Esteban-Puyuelo and B. Sanyal, *Role of Defects in Ultrafast Charge Recombination in Monolayer MoS 2*, Phys. Rev. B **103**, 235433 (2021).

[14] S. Tongay, J. Suh, C. Ataca, W. Fan, A. Luce, J. S. Kang, J. Liu, C. Ko, R. Raghunathanan, and J. Zhou, *Defects Activated Photoluminescence in Two-Dimensional Semiconductors: Interplay between Bound, Charged and Free Excitons*, Sci. Rep. **3**, 2657 (2013).

[15] S. Roy, W. Choi, S. Jeon, D.-H. Kim, H. Kim, S. J. Yun, Y. Lee, J. Lee, Y.-M. Kim, and J. Kim, *Atomic Observation of Filling Vacancies in Monolayer Transition Metal Sulfides by Chemically Sourced Sulfur Atoms*, Nano Lett. **18**, 4523 (2018).

[16] D. Rhodes, S. H. Chae, R. Ribeiro-Palau, and J. Hone, *Disorder in van Der Waals Heterostructures of 2D Materials*, Nat. Mater. **18**, 541 (2019).

[17] F. Wu, A. Galatas, R. Sundararaman, D. Rocca, and Y. Ping, *First-Principles Engineering of Charged Defects for Two-Dimensional Quantum Technologies*, Phys. Rev. Mater. **1**, 71001 (2017).

[18] S. Baek, E.-C. Shin, S. Cho, H.-K. Lyeo, and Y.-H. Kim, *Nanoscale Homojunction Thermoelectric Generator Built in Defect-Engineered MoS 2*, Phys. Rev. B **107**, 195411 (2023).

[19] K. Chen, R. Ghosh, X. Meng, A. Roy, J.-S. Kim, F. He, S. C. Mason, X. Xu, J.-F. Lin, and D. Akinwande, *Experimental Evidence of Exciton Capture by Mid-Gap Defects in CVD Grown Monolayer MoSe2*, Npj 2D Mater. Appl. **1**, 15 (2017).

[20] V. Carozo, Y. Wang, K. Fujisawa, B. R. Carvalho, A. McCreary, S. Feng, Z. Lin, C. Zhou, N. Perea-López, and A. L. Elías, *Optical Identification of Sulfur Vacancies: Bound Excitons at the Edges of Monolayer Tungsten Disulfide*, Sci. Adv. **3**, e1602813 (2017).

[21] M. S. Kim, S. J. Yun, Y. Lee, C. Seo, G. H. Han, K. K. Kim, Y. H. Lee, and J. Kim, *Biexciton Emission from Edges and Grain Boundaries of Triangular WS2 Monolayers*, ACS Nano **10**, 2399 (2016).

[22] L. Wang, I. Meric, P. Y. Huang, Q. Gao, Y. Gao, H. Tran, T. Taniguchi, K. Watanabe, L. M. Campos, and D. A. Muller, *One-Dimensional Electrical Contact to a Two-Dimensional Material*, Science (80-. ). **342**, 614 (2013).

[23] M.-L. Lin, Y. Zhou, J.-B. Wu, X. Cong, X.-L. Liu, J. Zhang, H. Li, W. Yao, and P.-H. Tan, *Cross-Dimensional Electron-Phonon Coupling in van Der Waals Heterostructures*, Nat. Commun. **10**, 2419 (2019).

[24] C. Jin, J. Kim, J. Suh, Z. Shi, B. Chen, X. Fan, M. Kam, K. Watanabe, T. Taniguchi, and S. Tongay, *Interlayer Electron–Phonon Coupling in WSe2/HBN Heterostructures*, Nat. Phys. **13**, 127 (2017).

[25] G. J. Slotman, G. A. de Wijs, A. Fasolino, and M. I. Katsnelson, *Phonons and Electron-phonon Coupling in Graphene-h-BN Heterostructures*, Ann. Phys. **526**, 381 (2014).

[26] P. Merkl, C.-K. Yong, M. Liebich, I. Hofmeister, G. Berghäuser, E. Malic, and R. Huber, *Proximity Control of Interlayer Exciton-Phonon Hybridization in van Der Waals*

*Heterostructures*, Nat. Commun. **12**, 1719 (2021).

[27] Q. Qian, Z. Zhang, and K. J. Chen, *Layer-Dependent Second-Order Raman Intensity of MoS 2 and WS e 2: Influence of Intervalley Scattering*, Phys. Rev. B **97**, 165409 (2018).

[28] S. Mignuzzi, A. J. Pollard, N. Bonini, B. Brennan, I. S. Gilmore, M. A. Pimenta, D. Richards, and D. Roy, *Effect of Disorder on Raman Scattering of Single-Layer Mo S 2*, Phys. Rev. B **91**, 195411 (2015).

[29] J. Yoo, K. Yang, B. W. Cho, K. K. Kim, S. C. Lim, S. M. Lee, and M. S. Jeong, *Identifying the Origin of Defect-Induced Raman Mode in WS2 Monolayers via Density Functional Perturbation Theory*, J. Phys. Chem. C **126**, 4182 (2022).

[30] C. Lee, B. G. Jeong, S. J. Yun, Y. H. Lee, S. M. Lee, and M. S. Jeong, *Unveiling Defect-Related Raman Mode of Monolayer WS2 via Tip-Enhanced Resonance Raman Scattering*, ACS Nano **12**, 9982 (2018).

[31] J. Ji, A. Zhang, J. Fan, Y. Li, X. Wang, J. Zhang, E. W. Plummer, and Q. Zhang, *Giant Magneto-Optical Raman Effect in a Layered Transition Metal Compound*, Proc. Natl. Acad. Sci. **113**, 2349 (2016).

[32] Y. Wan, X. Cheng, Y. Li, Y. Wang, Y. Du, Y. Zhao, B. Peng, L. Dai, and E. Kan, *Manipulating the Raman Scattering Rotation via Magnetic Field in an MoS 2 Monolayer*, RSC Adv. **11**, 4035 (2021).

[33] Y. Yang, W. Liu, Z. Lin, K. Zhu, S. Tian, Y. Huang, C. Gu, and J. Li, *Micro-Defects in Monolayer MoS2 Studied by Low-Temperature Magneto-Raman Mapping*, J. Phys. Chem. C **124**, 17418 (2020).

[34] A. Berkdemir, H. R. Gutiérrez, A. R. Botello-Méndez, N. Perea-López, A. L. Elías, C.-I. Chia, B. Wang, V. H. Crespi, F. López-Urías, and J.-C. Charlier, *Identification of Individual and Few Layers of WS2 Using Raman Spectroscopy*, Sci. Rep. **3**, 1755 (2013).

[35] N. Peimyoo, J. Shang, W. Yang, Y. Wang, C. Cong, and T. Yu, *Thermal Conductivity Determination of Suspended Mono-and Bilayer WS 2 by Raman Spectroscopy*, Nano Res. **8**, 1210 (2015).

[36] W. Zhao, Z. Ghorannevis, K. K. Amara, J. R. Pang, M. Toh, X. Zhang, C. Kloc, P. H. Tan, and G. Eda, *Lattice Dynamics in Mono-and Few-Layer Sheets of WS 2 and WSe 2*, Nanoscale **5**, 9677 (2013).

[37] S. Reich, A. C. Ferrari, R. Arenal, A. Loiseau, I. Bello, and J. Robertson, *Resonant Raman Scattering in Cubic and Hexagonal Boron Nitride*, Phys. Rev. B **71**, 205201 (2005).

[38] R. V Gorbachev, I. Riaz, R. R. Nair, R. Jalil, L. Britnell, B. D. Belle, E. W. Hill, K. S. Novoselov, K. Watanabe, and T. Taniguchi, *Hunting for Monolayer Boron Nitride: Optical and Raman Signatures*, Small **7**, 465 (2011).

[39] *Supplemental Material*.

[40] J. Park, M. S. Kim, E. Cha, J. Kim, and W. Choi, *Synthesis of Uniform Single Layer WS2 for Tunable Photoluminescence*, Sci. Rep. **7**, 16121 (2017).

[41] H. Wang, C. Zhang, and F. Rana, *Ultrafast Dynamics of Defect-Assisted Electron–Hole Recombination in Monolayer MoS2*, Nano Lett. **15**, 339 (2015).

[42] G. Moody, K. Tran, X. Lu, T. Autry, J. M. Fraser, R. P. Mirin, L. Yang, X. Li, and K. L. Silverman, *Microsecond Valley Lifetime of Defect-Bound Excitons in Monolayer WSe 2*, Phys. Rev. Lett. **121**, 57403 (2018).

[43] N. Saigal and S. Ghosh, *Evidence for Two Distinct Defect Related Luminescence Features in Monolayer MoS2*, Appl. Phys. Lett. **109**, 122105 (2016).

[44] E. Mitterreiter, B. Schuler, A. Micevic, D. Hernangómez-Pérez, K. Barthelmi, K. A. Cochrane, J. Kiemle, F. Sigger, J. Klein, and E. Wong, *The Role of Chalcogen Vacancies for Atomic Defect Emission in MoS2*, Nat. Commun. **12**, 3822 (2021).

[45] H. Qiu, T. Xu, Z. L. Wang, W. Ren, H. Y. Nan, Z. H. Ni, Q. Chen, S. J. Yuan, F. Miao, and F. Q. Song, *Modulating the Properties of MoS2 by Plasma Thinning and Defect Engineering*, Nat. Commun **4**, 2642 (2013).

[46] S. Paul, S. Karak, A. Mathew, A. Ram, and S. Saha, *Electron-Phonon and Phonon-Phonon Anharmonic Interactions in 2 H− Mo X 2 (X= S, Te): A Comprehensive Resonant Raman Study*, Phys. Rev. B **104**, 75418 (2021).

[47] H. Zobeiri, S. Xu, Y. Yue, Q. Zhang, Y. Xie, and X. Wang, *Effect of Temperature on Raman Intensity of Nm-Thick WS 2: Combined Effects of Resonance Raman, Optical Properties, and Interface Optical Interference*, Nanoscale **12**, 6064 (2020).

[48] J. Yang, J.-U. Lee, and H. Cheong, *Excitation Energy Dependence of Raman Spectra of Few-Layer WS2*, FlatChem **3**, 64 (2017).

[49] Q.-H. Tan, Y.-J. Sun, X.-L. Liu, Y. Zhao, Q. Xiong, P.-H. Tan, and J. Zhang, *Observation of Forbidden Phonons, Fano Resonance and Dark Excitons by Resonance Raman Scattering in Few-Layer WS2*, 2D Mater. **4**, 31007 (2017).

[50] S. Bae, K. Matsumoto, H. Raebiger, K. Shudo, Y.-H. Kim, Ø. S. Handegård, T. Nagao, M. Kitajima, Y. Sakai, and X. Zhang, *K-Point Longitudinal Acoustic Phonons Are Responsible for Ultrafast Intervalley Scattering in Monolayer MoSe2*, Nat. Commun. **13**, 4279 (2022).

[51] J. Hong, Z. Hu, M. Probert, K. Li, D. Lv, X. Yang, L. Gu, N. Mao, Q. Feng, and L. Xie, *Exploring Atomic Defects in Molybdenum Disulphide Monolayers*, Nat. Commun. **6**, 6293 (2015).

[52] H. Y. Jeong, Y. Jin, S. J. Yun, J. Zhao, J. Baik, D. H. Keum, H. S. Lee, and Y. H. Lee, *Heterogeneous Defect Domains in Single-crystalline Hexagonal WS2*, Adv. Mater. **29**, 1605043 (2017).

[53] T. T. Tran, K. Bray, M. J. Ford, M. Toth, and I. Aharonovich, *Quantum Emission from Hexagonal Boron Nitride Monolayers*, Nat. Nanotechnol. **11**, 37 (2016).

[54] J.-H. Chen, C. Jang, S. Xiao, M. Ishigami, and M. S. Fuhrer, *Intrinsic and Extrinsic Performance Limits of Graphene Devices on SiO2*, Nat. Nanotechnol. **3**, 206 (2008).

[55] D. B. Farmer, H.-Y. Chiu, Y.-M. Lin, K. A. Jenkins, F. Xia, and P. Avouris, *Utilization of a Buffered Dielectric to Achieve High Field-Effect Carrier Mobility in Graphene Transistors*, Nano Lett. **9**, 4474 (2009).

# Tailoring the defects and electronic band structure in WS$_2$/h-BN heterostructure

Suvodeep Paul, Saheb Karak, Saswata Talukdar, Devesh Negi, and Surajit Saha*

*Department of Physics, Indian Institute of Science Education and Research Bhopal, Bhopal 462066, India.*

**Supplemental Material**

**Supplementary note 1:**

The Raman measurements on the WS$_2$-hBN heterostructure clearly show defect-scattered phonons, which we have studied in detail in the present work. Additionally, we observe through other light emission (photoluminescence) measurements that the WS$_2$-hBN heterostructure is also associated with certain defect-induced mid-gap electronic states. The introduction of these additional features in the phonon and the electronic band structures has been observed to be associated with the capping layer of h-BN and may have been induced by the cross-planar electron-phonon coupling (EPC) in the heterostructures at the interface. The optical images obtained through a microscope fitted with an objective lens 50 × magnification (as shown in Figure SF2) and atomic force microscopy measurements (also shown in Figure SF2) provide the evidence of WS$_2$-h-BN heterostructure, further evidence of the presence of the h-BN capping layer can be seen as the presence of the $E_{2g}$ type Raman mode near 1366 cm$^{-1}$ associated with the in-plane B-N vibrations, as observed in Figure SF1.

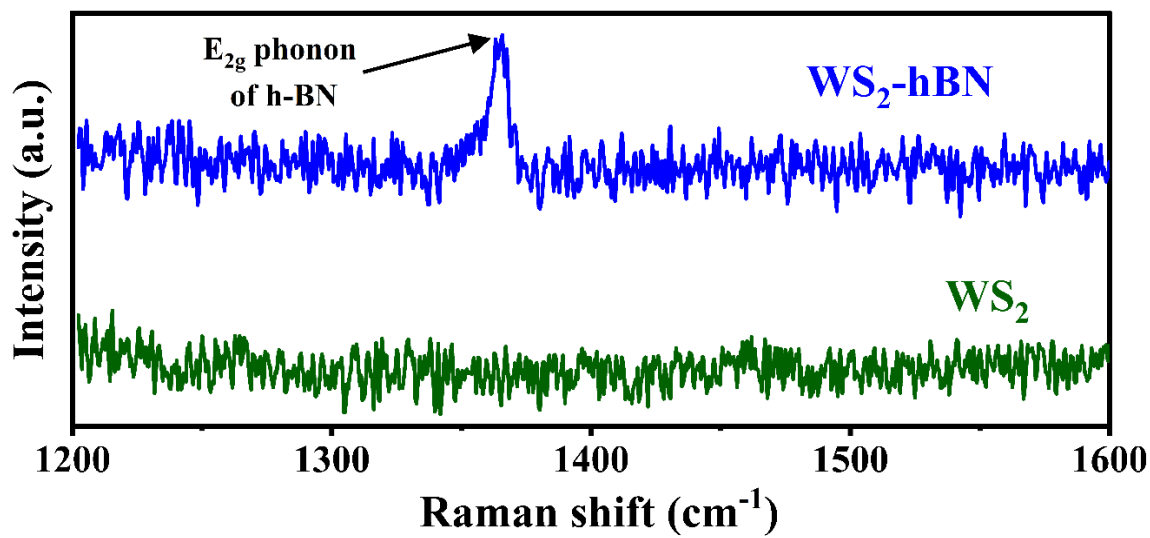

**Figure SF6.** Raman spectra of the uncapped WS$_2$ and the WS$_2$-hBN heterostructure, showing the h-BN $E_{2g}$ phonon in the heterostructure, confirming the presence of h-BN capping layer.

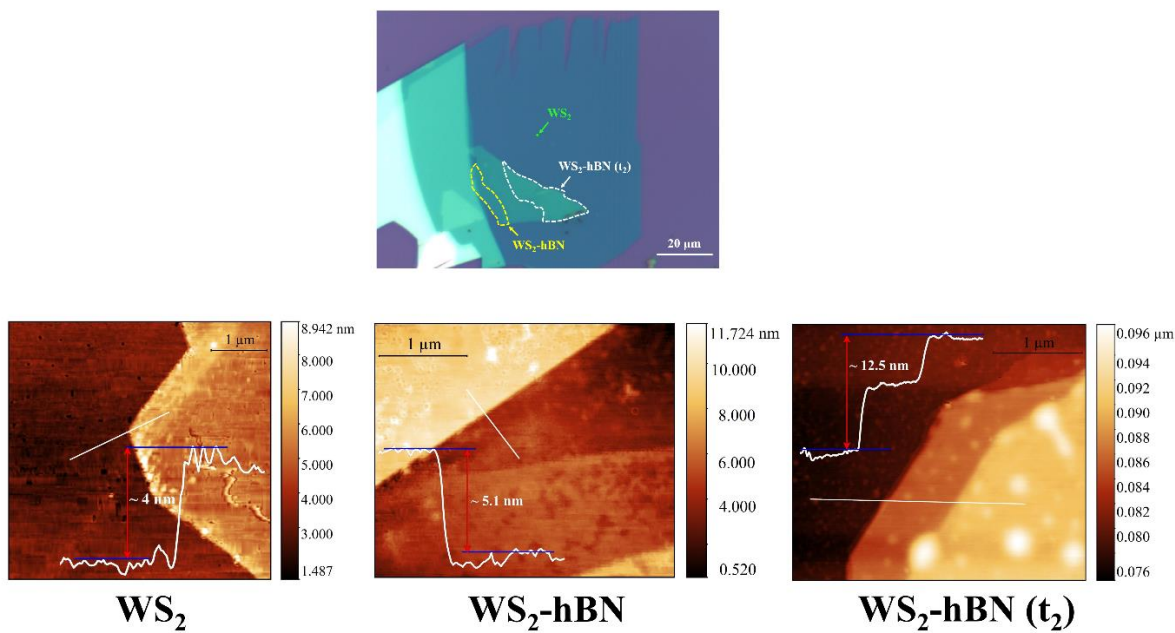

**Figure SF2.** The optical microscopic image of the various investigated samples and the corresponding AFM profiles used to measure the thicknesses of the flakes.

**Supplementary note 2:**

The Raman measurements on the WS$_2$-hBN heterostructures, as discussed in the main text constitute certain additional peaks, *viz.*, P$_1$ and P$_2$. Further, we observe an enhancement in the intensity of the P$_3$ mode, which is the $LA(M)$ phonon, associated with defect scattering in 2H transition metal dichalcogenides. Notably, the new modes P$_1$ and P$_2$ have not been previously reported. However, the appearance of these modes occurs simultaneously with the enhancement of the P$_3$ mode, giving us the impression that P$_3$ is also associated with defect scattering and is a combination mode of $LA(M)$ phonons. In order to confirm this proposition, we have also performed polarization-angle dependent Raman scattering measurements (Figure SF3) where we have clearly observed a similar polarization angle dependence for the $P_1$, $P_2$, and $P_3$ modes. While the modes $P_1$, $P_2$, and $P_3$ have not been previously observed or attributed to defect-scattering unambiguously for $WS_2$, certain theoretical and experimental works assigned the D mode to be related to sulfur vacancy defects in $WS_2$. However, the proximity of the D mode (~417 cm$^{-1}$) to the $P_{10}$ mode [$A_{1g}(\Gamma)$ phonon] and thermal broadening effects at room temperature makes it difficult to resolve the D mode through normal Raman scattering experiments. Therefore, we have recorded the Raman spectra of the uncapped $WS_2$ and the $WS_2$-hBN heterostructure using the 633 nm excitation line at 80 K (Figure SF4). Comparison of the two spectra clearly show an enhancement of the D mode in the $WS_2$-hBN heterostructure as also observed for the $P_3$ mode.

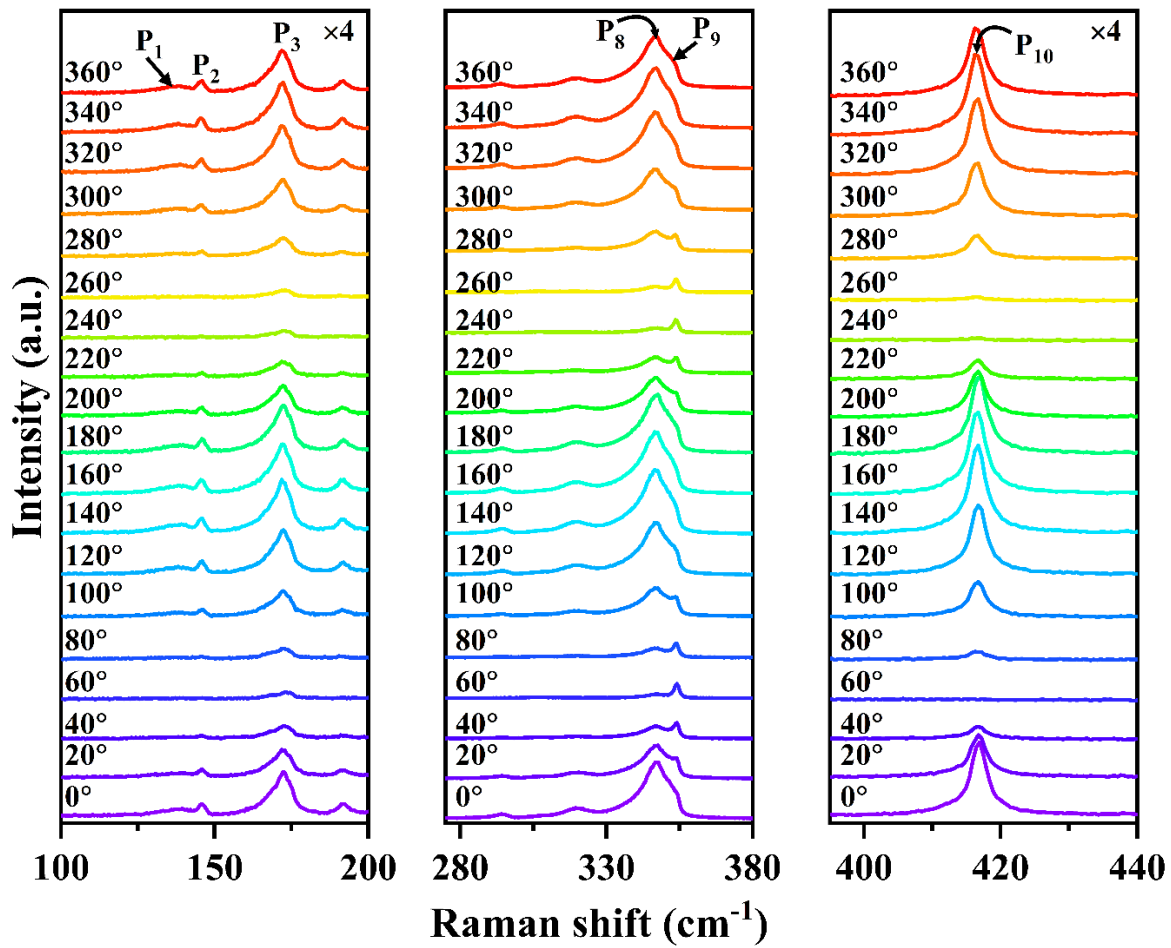

**Figure SF3.** Stack of polarization-angle dependent Raman spectra for the WS$_2$-hBN heterostructure. The prominent modes are labeled. The modes P$_1$-P$_3$ (left panel), P$_8$ (middle panel), and P$_{10}$ (right panel) show strong dependence on the polarization angle, while the mode P$_9$ (middle panel) shows no appreciable dependence.

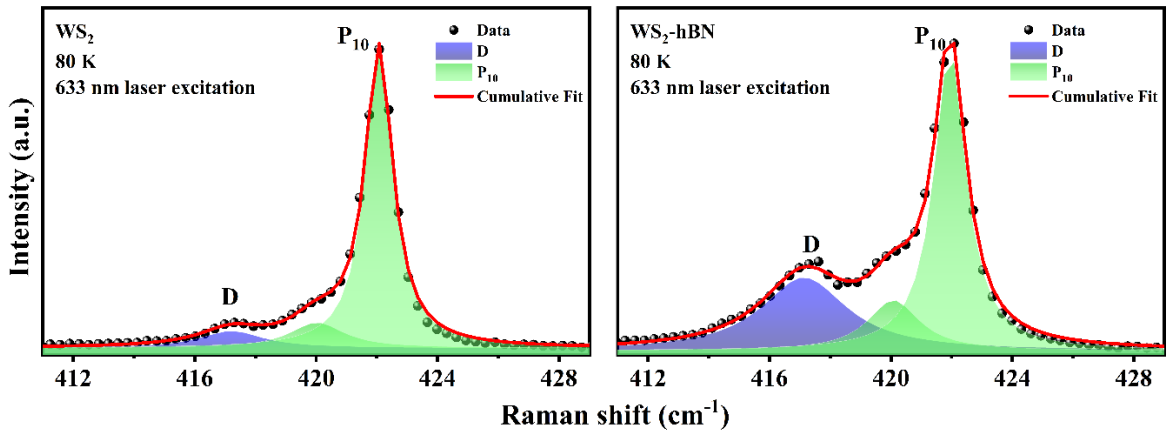

**Figure SF4.** Raman spectra of the uncapped $WS_2$ and the $WS_2$-hBN heterostructure obtained at 80 K temperature using 633 nm excitation laser line. The D peak (blue) at ~417 cm$^{-1}$ associated with S-vacancy defects in $WS_2$, clearly shows a strong enhancement in the $WS_2$-hBN heterostructure with respect to the uncapped $WS_2$.

**Supplementary note 3:**

The presence of defects can lead to the formation of mid-gap states in the electronic band structure of $WS_2$ below the exciton states. These mid-gap states may correspond to defect bound excitons and biexcitons [1]. In order to determine the exact nature of the states, we have performed laser power-dependent photoluminescence measurements. While the intensities of the excitons are supposed to show a linear dependence on the laser power, defect bound excitons and biexcitons show sublinear and superlinear dependencies. Our laser power dependent PL measurements (Figure SF5) clearly show features (corresponding to the defect-related mid-gap states) below the strong PL signatures corresponding to the A exciton and trion peaks, particularly at higher laser power. Figure SF6 shows the deconvolution of the PL spectra for the $WS_2$-hBN heterostructure obtained using 0.86 mW laser power into the constituent features using Gaussian profiles.

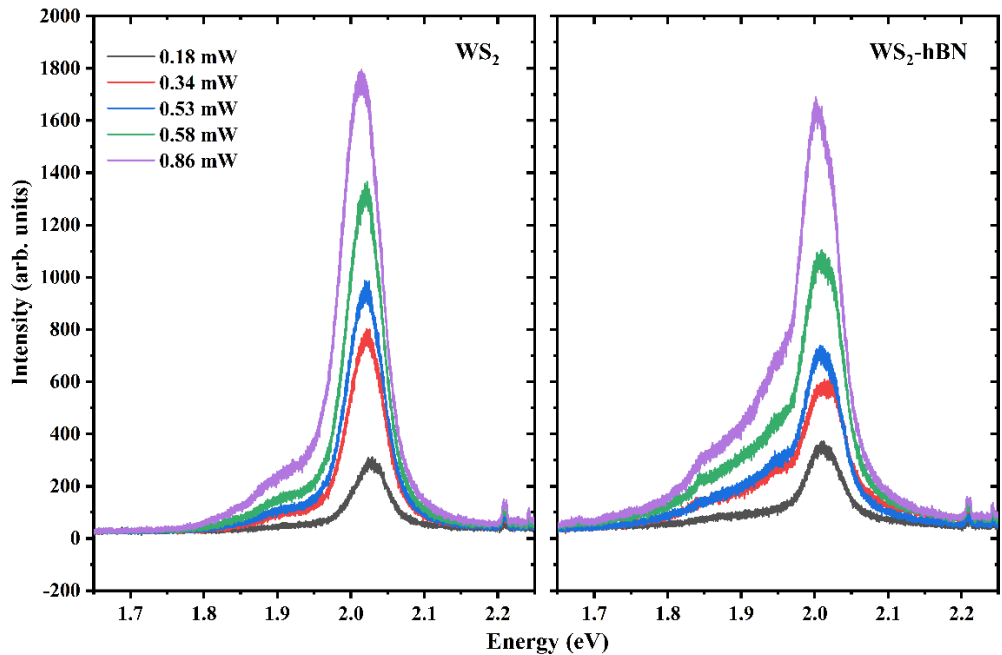

**Figure SF5.** Stack of photoluminescence spectra of the uncapped $WS_2$ and the $WS_2$-hBN heterostructure obtained at 80 K using different excitation laser powers.

**Supplementary note 4:**

The enhancement of Raman modes with temperature occurs by virtue of the resonance of the excitation laser with an electronic state in the electronic band structure. For $WS_2$, the electronic band gap is dominated by the influence of the A and B exciton states which lie below the band edge of the conduction band. The Raman scattering measurements performed using the 633 nm laser line show resonance with the A exciton bandgap, while any measurements performed using the 532 nm laser line show resonance with the B exciton bandgap. As discussed in the main text, the resonance phenomenon (observed through the intensity of the Raman modes) shows a dome-shaped dependence on temperature as a result of the decreasing excitonic band-gap with temperature (Figure SF7). This is explained with a schematic diagram in Figure 4 of the main text. The defect-scattered Raman modes, on the other hand, show resonance with the defect-related mid-gap electronic states, corresponding to the A exciton. The discussion in the main text mainly refers to the measurements performed with the 633 nm laser excitation. Similarly, we have also performed experiments with the 532 nm laser excitation, which resonates with the B exciton and shows a very similar dome-shaped dependence on temperature (Fig SF8). The maxima observed in the defect-scattered modes, in this case, correspond the resonance with defect related mid-gap states like defect-bound excitons and biexcitons associated with the B exciton.

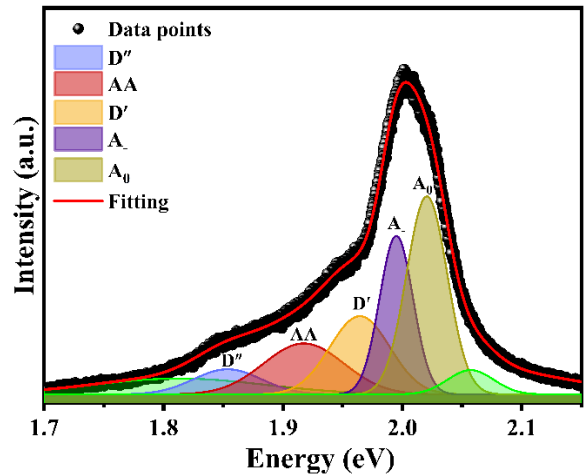

**Figure SF6.** The deconvolution (using Gaussian multi-peak fitting) of the PL spectra of the $WS_2$-hBN heterostructure, showing the contributions from the various emission states like the exciton ($A_0$), trion ($A_-$), defect-bound excitons (D′, D″), and biexciton (AA). The red line shows the cumulative fitting of the profile adding all the components.

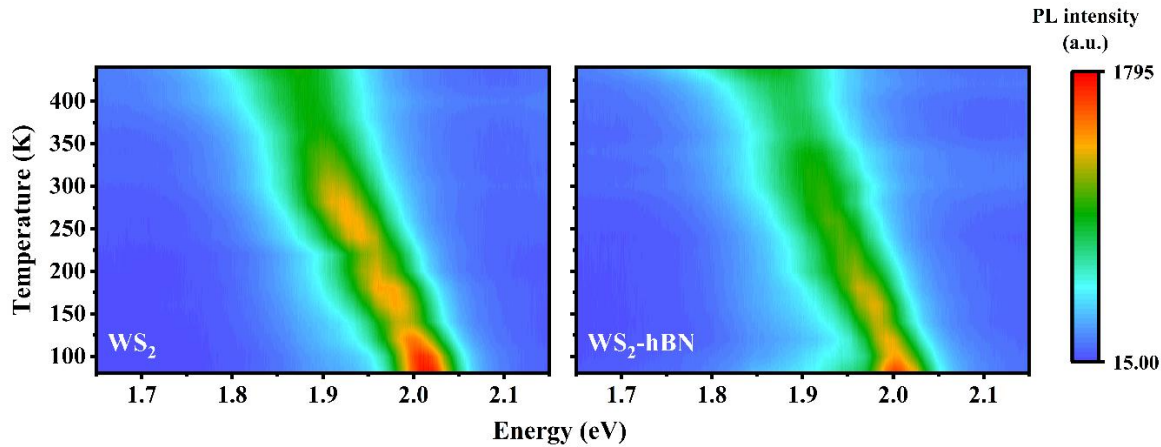

**Figure SF7.** The intensity profile of the photoluminescence spectra of the uncapped $WS_2$ and the $WS_2$-hBN heterostructure as a function of varying temperature, showing a redshift of the A exciton feature with increasing temperature.

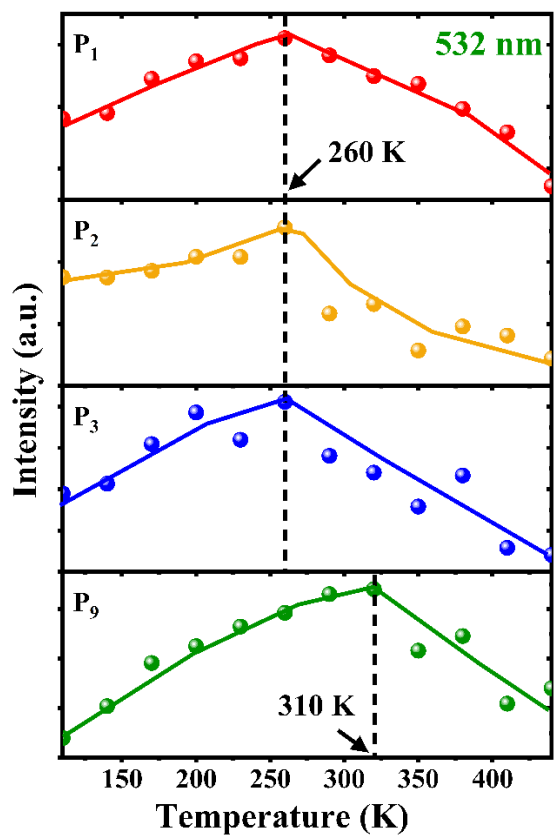

**Figure SF8.** The intensity of the defect-related modes ($P_1$, $P_2$, $P_3$) and the other prominent Raman mode ($P_9$) unrelated to defects clearly show the resonance phenomena at different temperatures, implying the involvement of different states in the resonance phenomena. These measurements were performed with the 532 nm (2.33 eV) laser excitation.

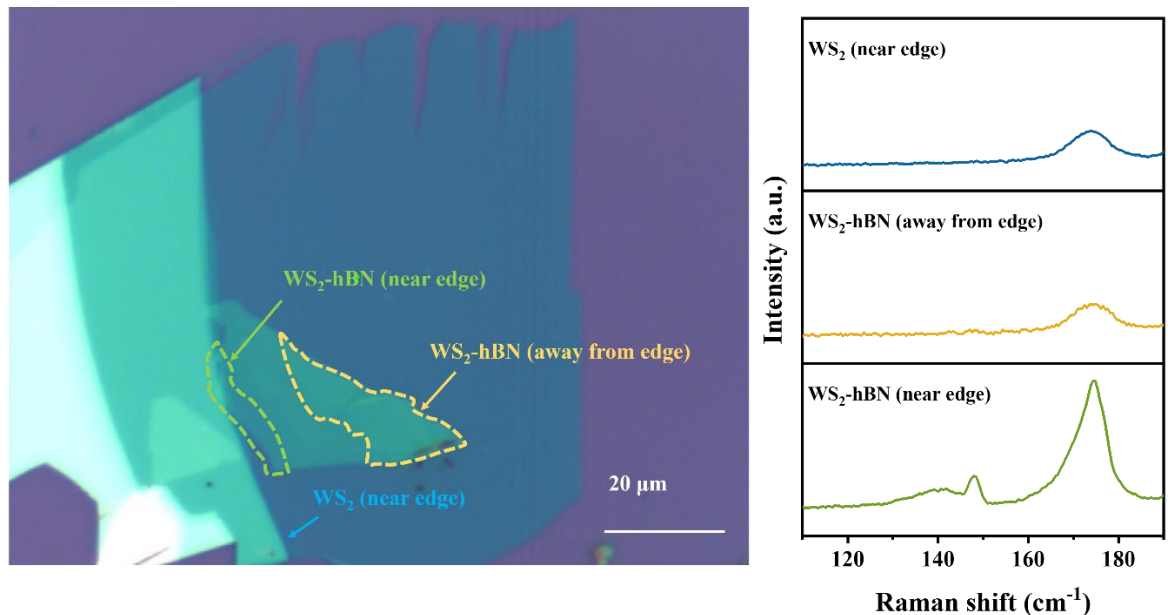

**Figure SF9.** Optical microscopy images and corresponding Raman spectra of the uncapped $WS_2$ (recorded near an edge) and two $WS_2$-h-BN heterostructures (one near the edge that was discussed in the main text and another located away from the edge) using the 532 nm excitation laser line. The spectra do not show any signatures of defect for the uncapped $WS_2$ (near edge) and the $WS_2$-hBN (away from edge) heterostructure.

**Supplementary note 5:**

To be noted that the signatures associated with the defects were observed in the $WS_2$-hBN heterostructure. The source of the defects was attributed to the proximity to an edge, which has been previously reported to show strong emission dominated by defect-bound excitons and biexcitons [1]. The flake edge as an origin of defect is further confirmed from the absence of any defect features in another $WS_2$-h-BN heterostructure based on the same $WS_2$ layer but at a location far away from the edges (Fig. SF9). Again, it was observed that the spectra of uncapped $WS_2$ recorded near the edge of the flake also show no appreciable signatures of the defects (Fig. SF9), indicating that along with the proximity to an edge, the presence of the capping layer of $WS_2$ further promotes the defect related phenomena. It has also been reported that the appearance of quasi-particle states like defect-bound excitons and biexcitons is also related to the availability of a large concentration of charge carriers [1]. We observe signatures of interlayer electron-phonon coupling (EPC) in the $WS_2$-h-BN heterostructure, which results in a localization of charge carriers in the $WS_2$ layer, thereby providing the required charge carriers for the creation of such quasiparticles. The presence of the interlayer coupling can be confirmed from the appearance of certain Raman-inactive interlayer phonons of h-BN (marked with black triangles in Fig. SF10) [2] and the enhancement of certain interlayer modes of WS$_2$.

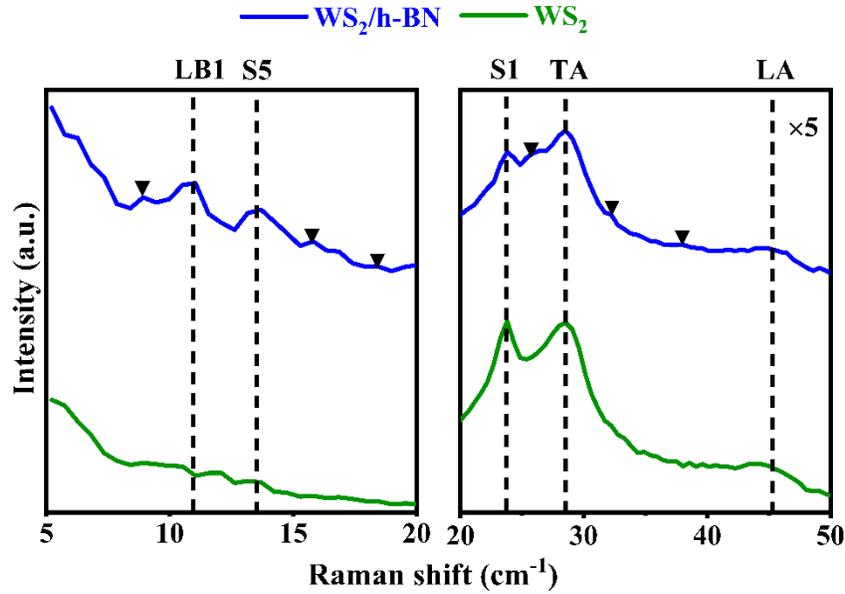

**Figure SF10.** The Raman-inactive interlayer phonon modes of h-BN (marked with black inverted triangles) observed in the spectrum of the WS$_2$-h-BN heterostructure because of the interlayer EPC. We further observe an enhancement of the LB1 and S5 interlayer phonons of WS$_2$ as well.

**References**

[1] M. S. Kim, S. J. Yun, Y. Lee, C. Seo, G. H. Han, K. K. Kim, Y. H. Lee, and J. Kim, *Biexciton Emission from Edges and Grain Boundaries of Triangular WS2 Monolayers*, ACS Nano **10**, 2399 (2016).

[2] M.-L. Lin, Y. Zhou, J.-B. Wu, X. Cong, X.-L. Liu, J. Zhang, H. Li, W. Yao, and P.-H. Tan, *Cross-Dimensional Electron-Phonon Coupling in van Der Waals Heterostructures*, Nat. Commun. **10**, 2419 (2019).